\documentclass[12pt,showpacs,showkeys,amsmath,aps,prc]{revtex4}

\usepackage{amsfonts} 
\usepackage{mathrsfs}
\usepackage{subfigure}
\usepackage{supertabular}
\usepackage{color,graphicx}
\usepackage[bookmarksopen]{hyperref}

\def  \p    {\pi}

\def  \th   {\theta}

\def  \veps {\varepsilon}

\def  \del  {\partial}

\def  \bef  {\begin{figure}}
\def  \eef  {\end{figure}}
\def  \be   {\begin{equation}}
\def  \ee   {\end{equation}}
\def  \ba   {\begin{array}}
\def  \ea   {\end{array}}
\def  \bea  {\begin{eqnarray}}
\def  \eea  {\end{eqnarray}}
\def  \beq  {\begin{eqnarray}}
\def  \eeq  {\end{eqnarray}}
\def  \nn   {\nonumber}
\def  \bd   {\begin{displaymath}}
\def  \ed   {\end{displaymath}}
\def  \bse  {\begin{subequations}}
\def  \ese  {\end{subequations}}
\def  \bwt  {\begin{widetext}}
\def  \ewt  {\end{widetext}}

\def  \ba   {{\bf{a_1}}}

\begin{document}
\title{Symmetric and anti-symmetric Landau parameters and 
magnetic properties of dense quark matter }
\author {Kausik Pal}
\author {Abhee K. Dutt-Mazumder}
\affiliation {High Energy Physics Division, Saha Institute of Nuclear Physics,
 1/AF Bidhannagar, Kolkata 700064, India.}

\medskip

\begin{abstract}
We calculate the dimensionless Fermi liquid parameters (FLPs), 
$F_{0,1}^{sym}$ and $F_{0,1}^{asym}$, for spin asymmetric 
dense quark matter. In general, the FLPs are infrared divergent
due to the exchange of massless gluons. To remove such divergences, 
the Hard Density Loop (HDL) corrected gluon propagator is used. 
The FLPs so determined are then invoked to calculate magnetic properties 
such as magnetization 
$\langle M\rangle$ and magnetic susceptibility $\chi_M$ 
of spin polarized quark matter. Finally, we investigate the 
possibility of magnetic instability by studying the density 
dependence of $\langle M\rangle$ and $\chi_M$.
\end{abstract}
\vspace{0.08 cm}



\maketitle

\section{Introduction}

The study of strongly interacting matter has been an area of 
contemporary research for quite sometime now. Such studies are usually 
made in the extreme condition of temperature and/or density. The high
temperature ($T$) studies are more relevant to the ultra-relativistic heavy ion 
collisions while the investigations involving high chemical potential ($\mu$)
or extreme case of cold matter are more appropriate to 
astrophysics \cite{baym_fri}.
It should, however, be noted that efforts are being directed recently
to also study the properties of a very dense system in the laboratory 
where matter with predominantly large chemical potential might be 
formed \cite{tahir_senger}.  We here restrict ourselves to zero 
temperature and investigate some of the properties of quark
matter in presence of a weak magnetic field.

It has been shown recently that the degenerate quark 
matter can show para-ferro phase transition below a critical density
\cite{tatsumi00}. To examine this possibility, in Ref.\cite{tatsumi00},
a variational calculation was performed. Subsequently, various
other calculations were also performed in different formalisms
to investigate such a possibility  with varied conclusions
\cite{niegawa05,nakano03,tatsumi01,son08,tatsumi06,ohnishi07,pal_gse,pal_sus}. 

The issue of spontaneous phase transition in dense
quark system at zero temperature
was also examined in \cite{pal09} by invoking Relativistic 
Fermi Liquid Theory (RFLT). In  particular, this was accomplished
by calculating the chemical potential ($\mu$) and energy density
of degenerate quark matter in terms of the Landau Parameters (LPs).
The RFLT was first developed by Baym and Chin 
\cite{baym76,baym_book} to study the properties of high density
nuclear matter. However, the formalism developed in Ref.\cite{baym76} 
is valid for unpolarized matter and 
LPs calculated there are spin averaged. Here, on the other
hand, we deal with polarized quark matter which requires
evaluation of the LPs with explicit spin dependencies.

Recently, in \cite{tatsumi08,tatsumi09} the authors have studied the 
magnetic properties of degenerate quark matter in presence of 
weak uniform external magnetic field $B$. Similar investigation
was also made in Ref.\cite{niegawa05} by evaluating the effective 
potential and employing quark magnetic moment as an order parameter. These 
calculations were, however, restricted to the case of unpolarized matter. 
On the contrary, our concern here is the magnetic properties of polarized quark
system. Consequently, we first determine various spin combination of LPs 
such as spin symmetric $(F_{0,1}^{+(-),sym})$ and spin anti-symmetric 
$(F_{0,1}^{+(-),asym})$ parameters and express quantities like magnetization 
and magnetic susceptibility 
in terms of these parameters. It is needless to mention
that unlike \cite{niegawa05,tatsumi08,tatsumi09}, the expressions for $\chi_M$ 
and $\langle M \rangle$, as presented here, depend on the spin polarization 
parameter $\xi=(n_q^{+}-n_q^{-})/(n_q^{+}+n_q^{-})$, where $n_q^+$ and
$n_q^-$ correspond to densities of spin-up and spin-down quarks, respectively.

It is well known that the calculations of LPs require evaluation of the
forward scattering amplitudes which are plagued with infrared divergences 
arising out of the exchange of massless gluons. Formally, such divergences 
can be removed by using HDL corrected gluon propagator. 
This can also be achieved by introducing screening mass for the gluons. Such
regularizations are necessary for the evaluation of individual LPs. On the
other hand, in various physical quantities like the ones we calculate here,
the LPs appear in particular combinations where such divergences cancel at
least to the order with which we are presently concerned.

The plan of the article is as follows. In Sec. II we derive 
the expressions of LPs for polarized quark matter. In Sec III, we 
calculate magnetic susceptibility in terms of LPs with 
explicit spin dependencies both with bare and HDL corrected 
gluon propagator. In Sec. IV we summarize and conclude.

\section{Symmetric and anti-symmetric Landau parameters }

In this section we calculate LPs for spin polarized quark matter.
We are dealing with quasi-particles whose spins are all
eigenstates of the spin along a given direction {\em viz.} $z$.
The quasiparticle interaction can be written as the sum of 
two parts {\em viz.} spin symmetric ($f_{pp'}^{sym}$) and 
anti-symmetric ($f_{pp'}^{asym}$) parameters \cite{baym_book,tatsumi09}:

\beq
f_{pp'}^{ss'} &=& f_{pp'}^{sym}+(s\cdot s') f_{pp'}^{asym}.
\eeq

Assuming that the spins are randomly oriented with respect to the momentum,
we take average over the angles $\theta_1$ and $\theta_2$ corresponding 
to spins $s$ and $s'$. The angular averaged interaction parameter 
is given by \cite{pal09}: 

\beq\label{ang_av_pp}
{\overline{f^{ss'}}}_{pp'}\Big|_{p=p_f^{s},p'=p_f^{s'}}&=&
\int \frac{\rm d\Omega_{1}}{4\pi}\int \frac{\rm d\Omega_{2}}{4\pi}
f^{ss'}_{pp'}\Big|_{p=p_f^{s},p'=p_f^{s'}}
\eeq

\footnotetext[1]{denoted hereafter as ${\overline{f_{pp'}^{ss'}}}= f_{pp'}^{ss'}$.}
$^1$ Here the spin may be either parallel 
($s=s'$) or anti-parallel ($s=-s'$) \cite{tatsumi00,pal09}. 
Thus the scattering possibilities are denoted by $(+,+)$, $(+,-)$, $(-,-)$ etc.
The interaction parameters can now be redefined as,

\beq
f_{pp'}^{++}&=&f_{pp'}^{sym}+f_{pp'}^{asym}~=~f_{pp'}^{--}\nn\\
f_{pp'}^{+-}&=&f_{pp'}^{sym}-f_{pp'}^{asym}~=~f_{pp'}^{-+}
\eeq

Once these interaction parameters are known, the FLPs can be determined
by expanding $f_{pp'}^{ss'}$ into the Legendre polynomial: 

\beq
f_l^{ss'}=(2l+1)\int\frac{d\Omega}{4\pi}P_{l}(\cos\theta)f_{pp'}^{ss'},
\eeq
where $\cos\theta={\hat p}\cdot{\hat p'}$.
We define symmetric and anti-symmetric part of 
LPs $f_l^{s,sym(asym)}$ what one does to
dealing with the isospins in nuclear matter \cite{baym_book,pal09}:

\beq
f^{+(-),sym}_{l}&=&\frac{1}{2}\Big(f^{++(--)}_{l}+f^{+-(-+)}_{l}\Big)\nn\\ 
f^{+(-),asym}_{l}&=&\frac{1}{2}\Big(f^{++(--)}_{l}-f^{+-(-+)}_{l}\Big)
\eeq
It should be noted here that, $f^{+-}_{pp'}=f^{-+}_{pp'}$. 

The dimensionless LPs are defined as 
$F_{l}^{s,sym(asym)}=N^{s}(0)f_{l}^{s,sym(asym)}$ 
\cite{pal09}, where $N^s(0)$ is the density of states at the 
Fermi surface, which can be written as,

\beq\label{dos} 
N^{s}(0)&=&\int\frac{\rm d^3{p}}{(2\pi)^3}
\delta(\veps_{ps}-\mu^{s})\nn\\
&=&\frac{N_c p_{f}^{s^2}}{2\pi^2}\left(\frac{\del p}{\del\veps_{ps}}
\right){\Big|}_{p=p_{f}^{s}}
\eeq
Here, $N_c$ is the color factor, $\veps_{ps}$ and $\mu^s$ are the spin 
dependent quasi-particle energy and chemical potential respectively.
It is evident from Eq.(\ref{dos}) that for spin
polarized matter, the density of states is spin dependent. This,
as we shall see, makes the calculation cumbersome. 
In the above expression $(\del p/\del\veps_{ps}){\Big|}_{p=p_{f}^{s}}$ 
is the inverse Fermi velocity $(1/v_{f}^{s})$, where $v_{f}^{s}$ is given by 
\cite{tatsumi08,pal09}

\beq
v_{f}^{s}&=& \frac{p_f^s}{\mu^s}
-\frac{N_c p_f^{s^2}}{2\pi^2}\frac{f_1^{s,sym}}{3}
\eeq
 
With the bare propagator, the angular averaged 
spin dependent interaction parameter yields \cite{pal09}

\beq
f^{++}_{pp'}{\Big|}_{p=p'=p_f^+}&=&-\frac{g^2}{9\veps_{f}^{+2}p_{f}^{+2}
(1-\cos\th)}\left[2m_{q}^2-p_{f}^{+2}(1-\cos\th)+\frac{2m_{q}p_{f}^{+2}}
{3(\veps_{f}^{+}+m_{q})}\right].\label{pp_int}\\
f^{+-}_{pp'}{\Big|}_{p=p_{f}^{+},p'=p_{f}^{-}}&=&
\frac{g^2}{9\veps_{f}^{+}\veps_{f}^{-}}
\left\{1-\left[\frac{m_{q}p_{f}^{+2}}{3(\veps_{f}^{+}+m_{q})}
+\frac{m_{q}p_{f}^{-2}}{3(\veps_{f}^{-}+m_{q})}\right]
\times\frac{1}{(m_{q}^{2}-\veps_{f}^{+}\veps_{f}^{-}
+p_{f}^{+}p_{f}^{-}\cos\th)}\right\}.\label{pm_int}\nn\\
\eeq
Here, $m_q$ is the quark mass, $p_{f}^{\pm}=p_{f}(1\pm{\xi})^{1/3}$,
$\veps_f^{\pm}=(p_f^{{\pm}^2}+m_q^2)^{1/2}$ and 
$p_f$ is the Fermi momentum of the unpolarized matter $({\xi}=0)$.
Similarly, $f_{pp'}^{--}$ can be obtained by replacing 
$p_f^+$ with $p_f^-$ and $\veps_f^+$ with $\veps_f^-$ in 
Eq.(\ref{pp_int}). One can find dimensionless LPs, $F_{0,1}^{sym}$ 
and $F_{0,1}^{asym}$ (suppressing spin indices) 
by considering OGE interaction. But both of these
$(F_{0,1}^{sym(asym)})$ exhibit infrared divergences because of 
the term $(1-\cos\theta)$ that appear in the denominator of 
the interaction parameter (see Eq.(\ref{pp_int})). This divergence 
disappears if one uses HDL corrected gluon propagator to evaluate the 
scattering amplitudes \cite{bel_kap}.

To construct HDL corrected gluon propagator with explicit spin dependence
one needs to evaluate the expressions for longitudinal $(\Pi_L)$ and
transverse $(\Pi_T)$ polarization which have been derived in \cite{pal_gse}. 
We borrow the results directly:

\beq
\Pi_{L}(k_0,k) &=&\frac{g^2}{4\p^2}(C_0^2-1)\sum_{s=\pm}
p_f^s\veps_f^s\left[-1+\frac{C_0}{2v_f^s}
\ln\left(\frac{C_0+v_f^s}{{C_0-v_f^s}}\right)\right],
\label{piL}\\
\Pi_{T}(k_0,k)&=&\frac{g^2}{16\p^2}C_0\sum_{s=\pm}{p_f^{s^2}}
\left[\frac{2C_0}{v_f^s}+
\left(1-\frac{C_0^2}{{v_f^s}^2}\right)
\ln\left(\frac{C_0+v_f^s}{{C_0-v_f^s}}\right)\right]
\label{piT}.
\eeq
Here, $C_0=k_0/|k|$, is the dimensionless variable and 
$v_f^{\pm}=p_f^{\pm}/\veps_f^{\pm}$. It might be noted
here, that the expressions for $\Pi_L$ and $\Pi_T$ look 
rather similar to what one obtains in the case of unpolarized 
matter ($\xi=0$)\cite{chin77} with only difference in $v_f^\pm$. 
In the static limit {\em i.e.} $C_0\rightarrow 0$, the spin 
dependent Debye mass $(m_D)$ is given by

\beq
\Pi_L&=&m_D^2~=~\frac{g^2}{4\p^2}\sum_{s=\pm}p_f^s\veps_f^s
\eeq
It is to be mentioned here, that the screening 
mass of the gluon is spin dependent and the transverse gluons
are screened only dynamically \cite{tatsumi08,tatsumi09}.
With these, the symmetric combination of dimensionless LPs are
found to be

\beq\label{F0_sym}
F_{0}^{+,sym}&=&\frac{g^2 p_f^{+}}{144\pi^2}
\Big\{\frac{1}{\veps_f^+}\Big[12-\frac{12m_q^3+12m_q^2\veps_f^{+}
+3m_q m_{D}^2+4p_f^{+2}m_q+3m_{D}^2\veps_f^{+}}{p_f^{+2}(m_q+\veps_f^+)}
\ln\Big(1+\frac{4p_f^{+2}}{m_D^2}\Big)\Big]\nn\\
&&+\frac{1}{\veps_f^-}\Big[12+\frac{1}{p_f^+ p_f^-(m_q+\veps_f^+)
(m_q+\veps_f^-)}\Big\{m_q^2[3m_D^2-2(p_f^{+2}+p_f^{-2})]\nn\\
&&+m_q[3m_D^2(\veps_f^{+}+\veps_f^-)
-2(\veps_f^{+}p_f^{-2}+\veps_f^{-}p_f^{+2})]+3m_D^2\veps_f^{+}\veps_f^-\Big\}
\nn\\&&\times
\ln\Big(\frac{2m_q^2-m_D^2+2p_f^+ p_f^{-}-2\veps_f^{+}\veps_f^-}
{2m_q^2-m_D^2-2p_f^+ p_f^{-}-2\veps_f^{+}\veps_f^-}\Big)\Big]\Big\}
\eeq

\beq\label{F1_sym}
F_{1}^{+,sym}&=&\frac{g^2 p_f^{+}}{48\pi^2}
\Big\{\frac{1}{\veps_f^+}\Big[\frac{12m_q^3+12m_q^2\veps_f^{+}
+3m_q m_{D}^2+4p_f^{+2}m_q+3m_{D}^2\veps_f^{+}}{p_f^{+2}(m_q+\veps_f^+)}
\nn\\&\times&
\Big[2-\Big(1+\frac{m_D^2}{2p_f^{+2}}\Big)
\ln\Big(1+\frac{4p_f^{+2}}{m_D^2}\Big)\Big]\Big]\nn\\&+&
\frac{1}{\veps_f^-}\Big[\frac{m_q^2[3m_D^2-2(p_f^{+2}+p_f^{-2})]
+m_q[3m_D^2(\veps_f^{+}+\veps_f^-)
-2(\veps_f^{+}p_f^{-2}+\veps_f^{-}p_f^{+2})]+3m_D^2\veps_f^{+}\veps_f^-}
{p_f^+ p_f^-(m_q+\veps_f^+)(m_q+\veps_f^-)}\Big]\nn\\&\times&
\Big[2-\Big(\frac{2m_q^2-m_D^2-2\veps_f^+\veps_f^-}{2p_f^+p_f^-}\Big)
\ln\Big(\frac{2m_q^2-m_D^2+2p_f^+ p_f^{-}-2\veps_f^{+}\veps_f^-}
{2m_q^2-m_D^2-2p_f^+ p_f^{-}-2\veps_f^{+}\veps_f^-}\Big)\Big]\Big]
\Big\}
\eeq


\begin{figure}[htb]
\begin{center}
\resizebox{7.5cm}{6.5cm}{\includegraphics[]{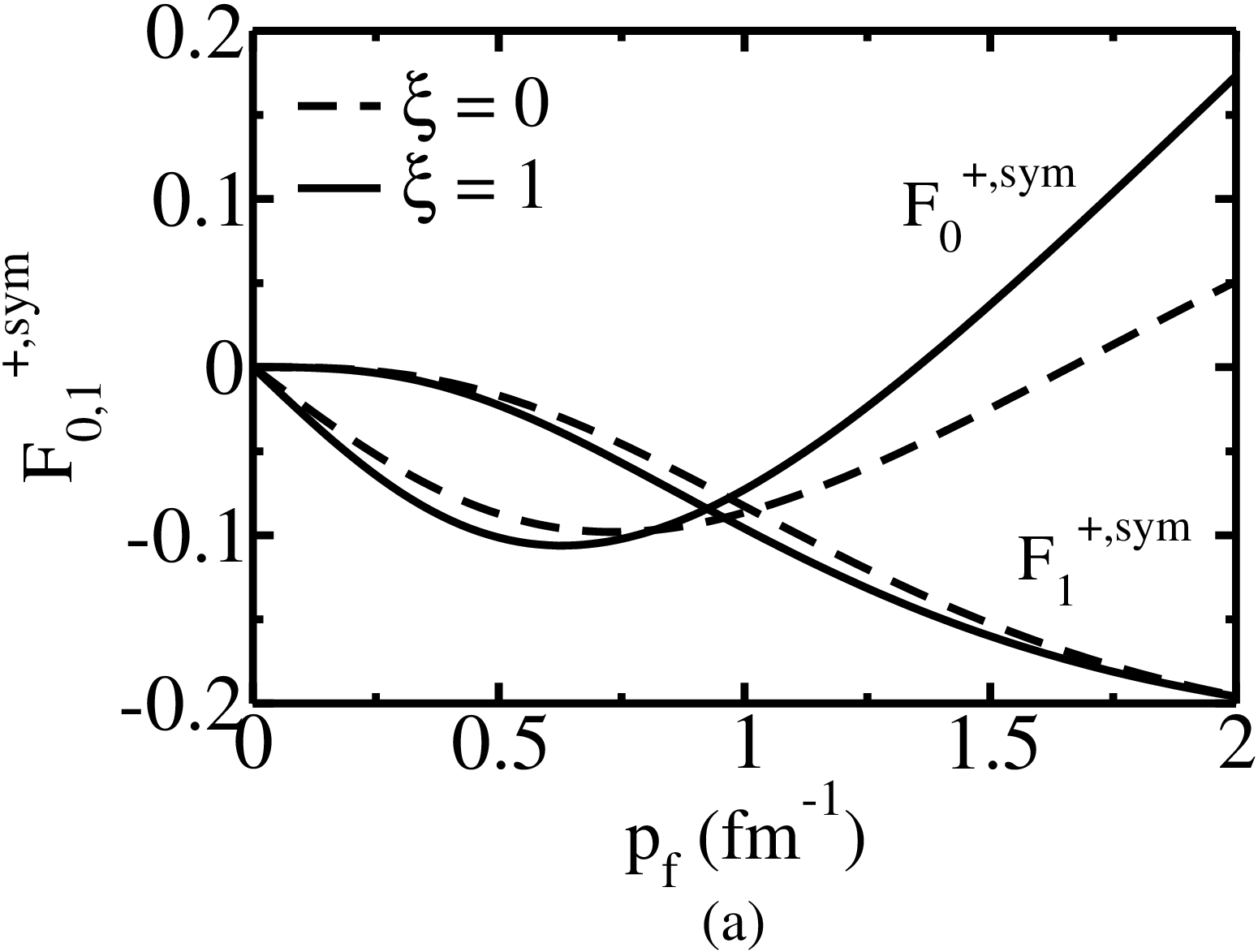}}~~~~
\resizebox{7.5cm}{6.5cm}{\includegraphics[]{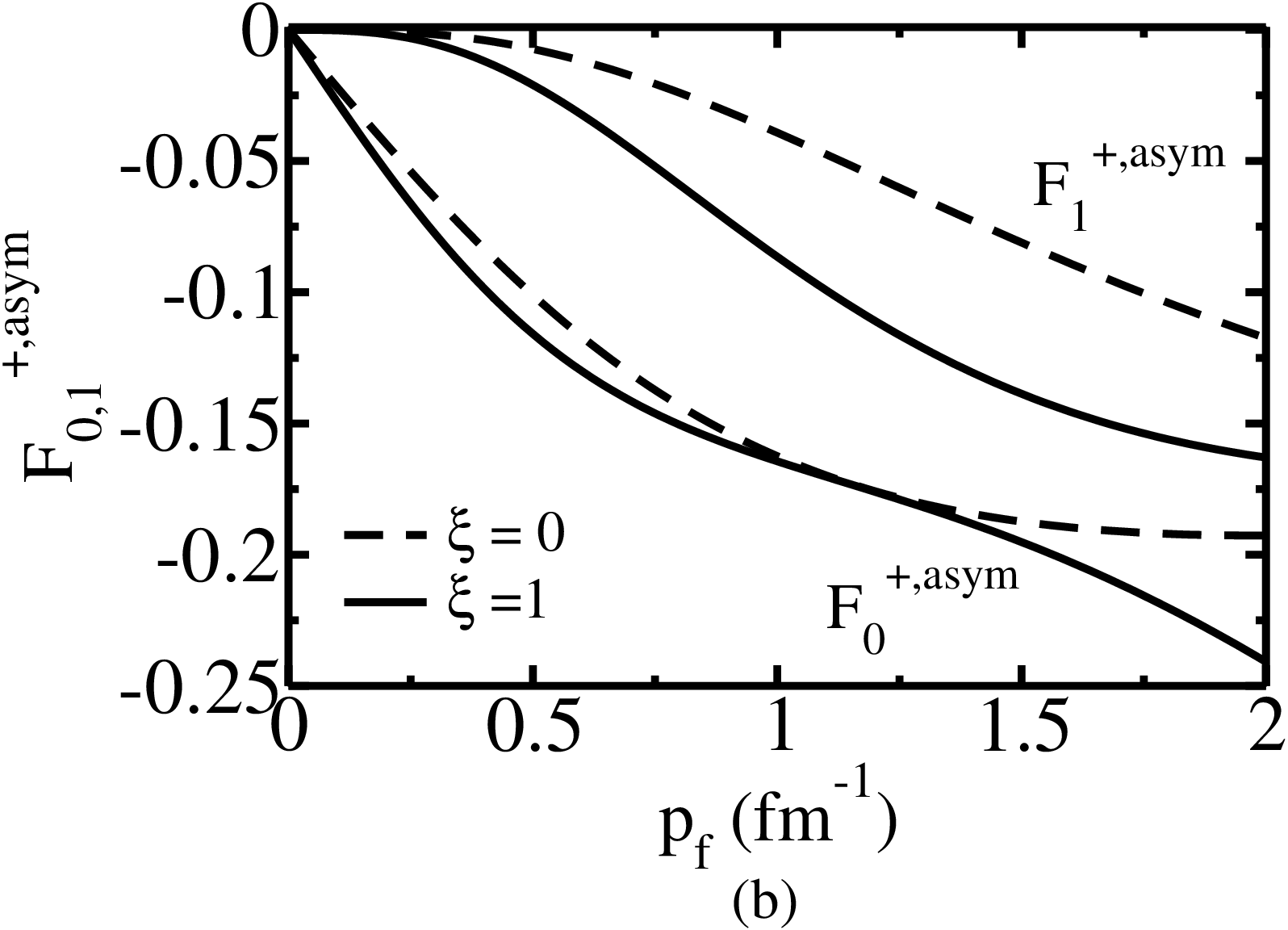}}
\caption{Dimensionless LPs as a function of Fermi momentum 
for unpolarized and polarized quark matter. 
Symmetric and anti-symmetric combination of LPs are plotted in 
$(a)$ and $(b)$ respectively.}
\label{flp}
\end{center}
\end{figure}


In deriving Eqs.(\ref{F0_sym}) and (\ref{F1_sym}), we consider exchange of 
longitudinal gluons only.
In Eqs.(\ref{F0_sym}) and (\ref{F1_sym}), the term in the first 
square bracket arises due to the scattering of like-spin states $(++)$, 
while the latter comes from the scattering of unlike-spin states $(+-)$. 
Similarly one may determine other combination of LPs like 
$F_{0,1}^{-,sym}$, $F_{0,1}^{+,asym}$, $F_{0,1}^{-,asym}$ etc. 
In Fig.{\ref{flp}}, density dependence of symmetric and anti-symmetric 
combination of dimensionless LPs is shown. Similar plots for the LPs in 
isospin asymmetric nuclear matter can be found in
\cite{aguirre07}. There, however, the calculated LPs are finite, as
the nucleon-nucleon interactions 
involve exchanges of massive mesons like 
$\sigma$, $\omega$, $\delta$ and $\rho$ etc. It is interesting to note
that the results of isospin asymmetric nuclear matter for the LPs
are qualitatively same as those of dense quark system.

\section {Magnetic susceptibility}

Now, we proceed to calculate the magnetic susceptibility for which
an uniform magnetic field $B$ is applied along the $z$ axis. The magnetic susceptibility is defined as \cite{tatsumi08,tatsumi09}

\beq\label{mag_sus01}
\chi_M &=& \sum_{f}\frac{\del \langle M\rangle_f}{\del B}{\Big|_{B=0}}
\eeq
where $\langle M\rangle_f$ is the magnetization for each flavor. 
Here, the magnetic field is considered to be
significantly weak for which the spinors remain unaffected and 
only modification enters through the single particle energy. 
Here, we consider one flavor quark matter and suppress the flavor indices.

In presence of constant magnetic field $B$, 
the magnetization depends on the difference of the number densities 
$\delta n_{p}^{asym}=\delta n_{p,s=1}-\delta n_{p,s=-1}$, where
\beq\label{nodensity}
n_{ps}=[1+exp~\beta(\veps_{ps}-\mu-\frac{1}{2}g_{D}(p)\mu_{q}sB)]^{-1}.
\eeq 
In the last equation, $\mu_q$ denotes the Dirac magnetic moment
and $g_{D}(p)$ is the gyromagnetic ratio. 
The magnetization is given by \cite{tatsumi08}

\beq\label{magnetiz}
\langle M\rangle &=& \frac{\mu_q}{2}N_c\int\frac{d^3 {\rm p}}
{(2\pi)^3}g_{D}(p)\delta n_{p}^{asym}~~.
\eeq

For constant magnetic field,
the variation of the distribution function yields \cite{tatsumi08,tatsumi09},

\beq
\delta n_{ps} &=& \frac{\delta n_{ps}}{\delta \veps_{ps}}
\left[-\frac{1}{2}g_D(p)\mu_{q}sB+N_{c}\sum_{s'}
\int\frac{d^3 {\rm p}}{(2\pi)^3}f_{pp'}^{ss'}\delta n_{p's'}\right]
\eeq
and $\delta n_{p}^{asym}$ is therefore given by

\beq\label{d_asym}
\delta n_{p}^{asym}&=&-\frac{1}{2}g_D(p)\mu_q B
\left(\frac{\del n_p^+}{\del\veps_p^+}
+\frac{\del n_p^-}{\del\veps_p^-}\right)
+N_c\frac{\del n_p^+}{\del\veps_p^+}
\left(f_0^{++}\delta n^{+}+f_0^{+-}\delta n^{-}\right)\nn\\
&&-N_c\frac{\del n_p^-}{\del\veps_p^-}
\left(f_0^{-+}\delta n^{+}+f_0^{--}\delta n^{-}\right)
\eeq

With the help of Eqs.(\ref{magnetiz}) and (\ref{d_asym}) the average
magnetization becomes,

\beq
\langle M\rangle &=& \frac{\frac{1}{4}{\overline{g}_D}^2\mu_q^2 B [N^+(0)+N^-(0)]} 
{1+[N^+(0)+N^-(0)]f_{0}^{asym}},
\eeq
where we have suppressed the spin indices for $f_l^{s,asym(sym)}$.
The expression of $\langle M \rangle$ may be compared with the one presented in 
\cite{tatsumi08,tatsumi09} to see the difference between the unpolarized
and polarized matter. Likewise, the magnetic susceptibility 
is found to be

\beq\label{mag_sus02}
\chi_{M}&=&\Big(\frac{\overline{g}_D\mu_q}{2}\Big)^2
\frac{[N^+(0)+N^-(0)]}{1+[N^+(0)+N^-(0)]f_{0}^{asym}}
\eeq
where $\overline{g}_D$ is the angular averaged gyromagnetic ratio
\cite{tatsumi08,tatsumi09}. 

With the help of Eqs.(\ref{dos}) and (\ref{mag_sus02}) we express the 
magnetic susceptibility in terms of LPs as,

\beq\label{mag_sus03}
\chi_M &=& \chi_P\Big[1+\frac{N_c(p_f^{+}\mu^{+}+p_f^{-}\mu^{-})}
{2\pi^2}\Big(f_0^{asym}-\frac{1}{3}f_1^{sym}\Big)\Big]^{-1}
\eeq
Here, $\chi_P={\overline g}_D^2\mu_q^2N_c(p_f^{+}\mu^{+}+p_f^{-}\mu^{-})/(8\pi^2)$
is the Pauli susceptibility \cite{tatsumi08,tatsumi09}. 
For unpolarized matter $\xi=0$, 
implying $p_f^{+}=p_f^{-}$, $\mu^{+}=\mu^{-}$ and $N^{+}(0)=N^{-}(0)$.
From Eq.(\ref{mag_sus03}) we get the well known result for 
magnetic susceptibility \cite{tatsumi08,tatsumi09}
\beq
\chi_M &=& \chi_P\Big[1+\frac{N_c p_f\mu}
{\pi^2}\Big(f_0^{asym}-\frac{1}{3}f_1^{sym}\Big)\Big]^{-1}
\eeq

\subsection {Susceptibility with bare propagator}

We have already mentioned that the individual LPs are infrared 
divergent when evaluated with the bare
gluon propagator. But the combination 
$\Big(f_0^{asym}-\frac{1}{3}f_1^{sym}\Big)$ 
is always finite and turns out to be

\beq\label{wos_para}
f_0^{asym}-\frac{1}{3}f_1^{sym}&=&\frac{1}{8}
\left[\int_{-1}^{+1}{d\rm}(\cos\theta)(1-\cos\theta)(f_{pp'}^{++}+f_{pp'}^{--})
\right.\nn\\&&\left.
-\int_{-1}^{+1}{\rm d}(\cos\theta)(1+\cos\theta)(f_{pp'}^{+-}+f_{pp'}^{-+})
\right]\nn\\
&=& I_1-I_2
\eeq

Using Eqs.(\ref{pp_int}), (\ref{pm_int}) and Eq.(\ref{wos_para}) we have

\beq
I_1 &=& -\frac{g^2}{36}\Big\{\frac{1}{p_f^{+2}\veps_f^{+2}}
\left[2m_{q}^2-p_{f}^{+2}+\frac{2m_{q}p_{f}^{+2}}
{3(\veps_{f}^{+}+m_{q})}\right]
+[p_f^{+}\rightarrow p_f^{-},\veps_f^{+}\rightarrow \veps_f^{-}]\Big\}
\label{wos_I1}\\
I_2 &=& \frac{g^2}{36\veps_f^+\veps_f^-}\times
\frac{1}{3p_f^{+2}p_f^{-2}(m_q+\veps_f^+)(m_q+\veps_f^-)}\nn\\
&\times&\Big\{-2p_f^+p_f^-\Big[p_f^-\veps_f^+(m_q p_f^{-}-3m_q p_f^+
-3p_f^+\veps_f^-)+m_q^2(p_f^{+2}-3p_f^+p_f^{-}+p_f^{-2})\nn\\
&&+m_qp_f^+\veps_f^-(p_f^+-3p_f^-)\Big]
+m_q\Big[\veps_f^+(m_q^2p_f^{-2}-p_f^+p_f^{-3}-p_f^{+2}\veps_f^{-2}
-m_q\veps_f^-[p_f^{+2}+p_f^{-2}])\nn\\
&&-p_f^{-2}\veps_f^-\veps_f^{+2}
+(m_q^2-p_f^+p_f^-)(\veps_f^-p_f^{+2}+m_q[p_f^{+2}+p_f^{-2}])\Big]
\ln\Big(\frac{m_q^2+p_f^+p_f^{-}-\veps_f^+\veps_f^-}
{m_q^2-p_f^+p_f^{-}-\veps_f^+\veps_f^-}\Big)\Big\}
\label{wos_I2}
\eeq
To determine $\chi_M$ for various $\xi$, we insert Eq.(\ref{wos_para})
in Eq.(\ref{mag_sus03}) where $I_1$ and $I_2$ are given by 
Eqs.(\ref{wos_I1}) and (\ref{wos_I2}).

\subsection {Susceptibility with HDL corrected propagator}

In this section we consider the screening effects 
due to HDL corrected propagator of the gauge field \cite{bel_kap}. 
The scattering amplitude can be written as \cite{tatsumi08}

\beq\label{amp_wm}
{\cal M}_{ps,p's'} &=& -\frac{4g^2}{9}
[{\cal T}^{00}(Ps,P's')D_{00}+{\cal T}^{ij}(Ps,P's')D_{ij}]
\eeq
In the coulomb gauge, we have $D_{00}=\Delta_l$ and
$D_{ij}=(\delta_{ij}-q_i q_j/q^2)\Delta_t$, where $q=p-p'$. 
$\Delta_l$ and $\Delta_t$ denote the longitudinal and transverse 
gluon propagators given by \cite{dutt_mazum05}

\beq
\Delta_l~=~\frac{1}{q^2+m_D^2}~,
&&\Delta_t~=~\frac{1}{q_0^2-q^2}
\eeq
For spin parallel $(s=s')$ and anti-parallel $(s=-s')$ interaction,
$\Delta_l$ and $\Delta_t$ have the following form :

\beq
{\Delta_l}(s=s')\Big|_{p=p'=p_f^{\pm}}&=&\frac{1}
{2p_f^{{\pm}^2}(1-\cos\th)+m_D^2},\nn\\
{\Delta_l}(s=-s')\Big|_{p=p_f^+,p'=p_f^-}&=&\frac{1}{p_f^{+2}+p_f^{-2}
-2p_f^{+}p_f^{-}(1-\cos\th)+m_D^2}.
\label{delta_l}\\
{\rm  ~and~~~~~~~}
{\Delta_t}(s=s')\Big|_{p=p'=p_f^{\pm}}&=&-\frac{1}
{2p_f^{{\pm}^2}(1-\cos\th)}~,\nn\\
{\Delta_t}(s=-s')\Big|_{p=p_f^+,p'=p_f^-}&=&\frac{1}
{2(m_q^2-\veps_f^{+}\veps_f^{-}+p_f^{+}p_f^{-}\cos\th)}
\label{delta_t}
\eeq

The matrix element given by Eq.(\ref{amp_wm}) can be calculated 
easily with OGE. We find that \cite{tatsumi09}
\beq
{\cal T}^{00}(Ps,P's')&=&{\rm Tr}[\gamma^{0}\rho(P,s)\gamma^{0}\rho(P',s')]\nn\\
&=&\frac{1}{4m_q^2}\Big[2p_0p'_0-P\cdot P'+m_q^2
+(m_q^2-P\cdot P')(2a_0b_0-a\cdot b)\nn\\
&&+2a_0p'_0(P\cdot b)-2p_0p'_0(a\cdot b)+2p_0b_0(a\cdot P')
-(P\cdot b)(P'\cdot a)\Big]
\eeq

and

\beq
{\cal T}^{ij}(Ps,P's')&=&{\rm Tr}[\gamma^{i}\rho(P,s)\gamma^{j}\rho(P',s')]\nn\\
&=&\frac{1}{4m_q^2}\Big\{(1-a\cdot b)\widehat{p^i{p'}^j}
+(m_q^2-P\cdot P')\widehat{a^ib^j}+(a\cdot P')\widehat{p^ib^j}\nn\\
&&+(b\cdot P)\widehat{{p'}^ia^j}+g^{ij}[(m_q^2-P\cdot P')(1-a\cdot b)
-(P\cdot b)(P'\cdot a)]\Big\},
\eeq
with a symbol $\widehat{a^ib^j}=a^ib^j+b^ia^j$. Here, the spin vector $a_{\mu}$ and  $b_{\mu}$ are given by

\beq
a~=~s+\frac{p(p\cdot s)}{m_{q}(\veps_{p}+m_{q})};~~
a^{0}~=~\frac{p\cdot s}{m_{q}}\\
b~=~s'+\frac{p'(p'\cdot s')}{m_{q}(\veps_{p'}+m_{q})};~~
b^{0}~=~\frac{p'\cdot s'}{m_{q}}
\eeq

To evaluate the spin symmetric and spin anti-symmetric combination of LPs,
we need to calculate the scattering amplitudes both for spin non-flip $(s=s')$ 
and spin flip $(s=-s')$ interactions. The traces relevant for the
longitudinal gluon exchange are given by,

\beq
{\cal T}_{00}^{++} &=& \frac{1}{6m_q^2(m_q+\veps_f^+)^2}
\Big[12m_q^4+12m_q^3\veps_f^{+}
+6m_q\veps_f^{+}p_f^{+2}(1+\cos\th)\nn\\
&&+6m_q^2p_f^{+2}(2+\cos\th)+p_f^{+4}(2+3\cos\th)\Big]
\label{M0pp}\\
{\cal T}_{00}^{+-} &=& \frac{p_f^{+2}p_f^{-2}}
{6m_q^2(m_q+\veps_f^+)(m_q+\veps_f^-)}
\label{M0pm}
\eeq

Similarly, the coefficient of $\Delta_t$ turns out to be

\beq
\Big[{\cal T}_{ij}\times(\delta^{ij}-\frac{q^iq^j}{q^2})\Big]^{++}
&=&-\frac{p_f^{+2}}{6m_q^3(m_q+\veps_f^+)^2}
\Big[6m_qp_f^{+2}+2p_f^{+2}\veps_f^{+}\nn\\
&&+2m_q^2\veps_f^{+}(4+3\cos\th)+m_q^3(8+3\cos\th)\Big]
\label{Mipp}
\eeq\\
\beq
\Big[{\cal T}_{ij}\times(\delta^{ij}-\frac{q^iq^j}{q^2})\Big]^{+-}
&=&-\frac{1}{6m_q^3(m_q+\veps_f^+)(m_q+\veps_f^-)
(p_f^{+2}+p_f^{-2}-2p_f^+p_f^-\cos\th)}\nn\\
&\times&\Big\{-p_f^{+2}[m_q(p_f^{+2}+p_f^{-2})+p_f^{-2}\veps_f^{+}
+p_f^{+2}\veps_f^{-}][2p_f^{-2}+m_q(m_q+\veps_f^-)]\nn\\
&&+m_q(m_q+\veps_f^+)\Big[-p_f^{-2}(m_q[p_f^{+2}+p_f^{-2}]
+p_f^{-2}\veps_f^{+}+p_f^{+2}\veps_f^{-})\nn\\
&&+2(p_f^{+2}-3p_f^{+}p_f^{-}+p_f^{-2})(m_q+\veps_f^-)
(m_q^2-\veps_f^+\veps_f^-)\Big]\nn\\
&&+p_f^+p_f^-\cos\th\Big[2p_f^{-2}(m_q[p_f^{+2}+p_f^{-2}]
+p_f^{-2}\veps_f^{+}+p_f^{+2}\veps_f^{-})\nn\\
&&+m_q(2m_q^2[2p_f^{+2}-3p_f^{+}p_f^{-}+2p_f^{-2}]
+p_f^{-2}\veps_f^{+2}+p_f^{+2}\veps_f^{-2}\nn\\
&&+m_q\veps_f^-[5p_f^{+2}-6p_f^{+}p_f^{-}+3p_f^{-2}]
+m_q\veps_f^+[3p_f^{+2}-6p_f^{+}p_f^{-}+5p_f^{-2}]\nn\\
&&+3\veps_f^+\veps_f^-[p_f^{+2}+p_f^{-2}-2p_f^+p_f^-\cos\th])\Big]\Big\}
\label{Mipm}
\eeq

Using Eqs.(\ref{amp_wm})-(\ref{delta_t}) and (\ref{M0pp})-(\ref{Mipm})
one can easily calculate the required combination 
$\Big(f_0^{asym}-\frac{1}{3}f_1^{sym}\Big)$ to evaluate the 
magnetic susceptibility. Inserting this particular combination of 
$f_0$ and $f_1$ in Eq.(\ref{mag_sus03}) we get $\chi_M$.
To determine $\chi_M$, we need to evaluate first $\mu^{+}$ and $\mu^{-}$.
This can be done by adopting the procedure outlined in Ref.\cite{pal09}.
With these, we can estimate $\chi_M$ numerically for the
polarized and unpolarized matter at various densities. The
corresponding results are discussed below.

\vskip 0.2in

\begin{figure}[htb]
\begin{center}
\resizebox{8.5cm}{7.0cm}{\includegraphics[]{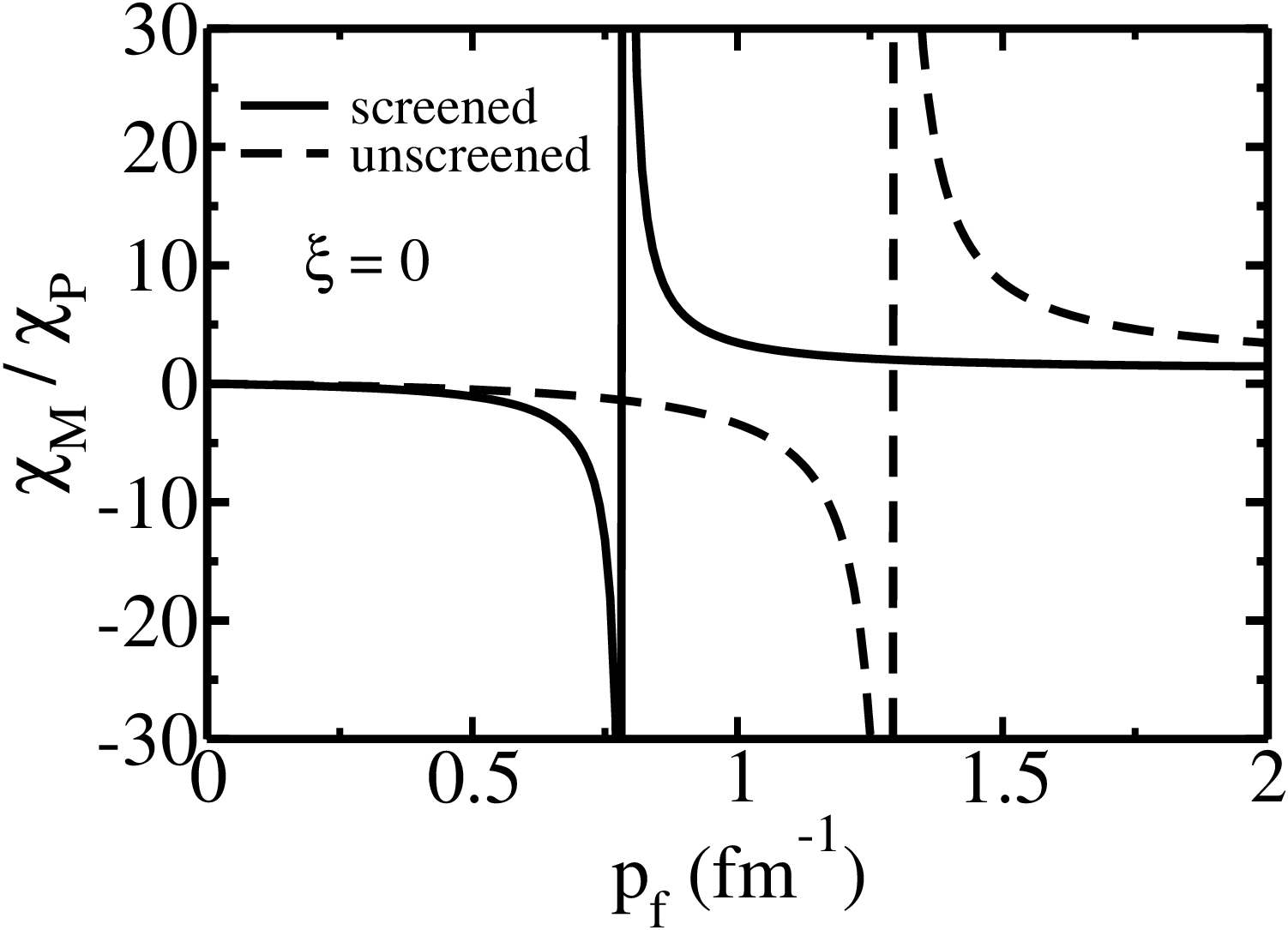}}
\caption{Density dependence of magnetic susceptibility. 
Screening effects (solid line) are compare 
with the simple OGE case (dashed line) for unpolarized quark matter.}
\label{chi_comp}
\end{center}
\end{figure}


In Fig.{\ref{chi_comp}} we plot the magnetic susceptibility of cold and dense 
unpolarized quark matter as a function of Fermi momentum. It is observed
that, upon inclusion of the screening effects, the divergence move towards 
lower densities. This is consistent with what one obtains for
unpolarized matter\cite{tatsumi08,tatsumi09}. Such shifts toward 
lower density are expected, as we know, that the screening effect 
weakens the Fock exchange interaction 
(See Ref.\cite{tatsumi08,tatsumi09}). 
Moreover, this divergence is related to the magnetic phase transition of quark 
matter which shows up when the square bracketed term in Eq.(\ref{mag_sus03})
vanishes. As noted in \cite{tatsumi00} and also  in \cite{pal09}, 
this density approximately corresponds to the critical density for
para-ferro phase transition. For the numerical estimation,
we take $\alpha_c=g^2/{4\pi}=2.2$ and $m_{q}=300 MeV$ 
\cite{tatsumi00,pal09,tatsumi08,tatsumi09}, 

\vskip 0.2in

\begin{figure}[htb]
\begin{center}
\resizebox{8.5cm}{7.0cm}{\includegraphics[]{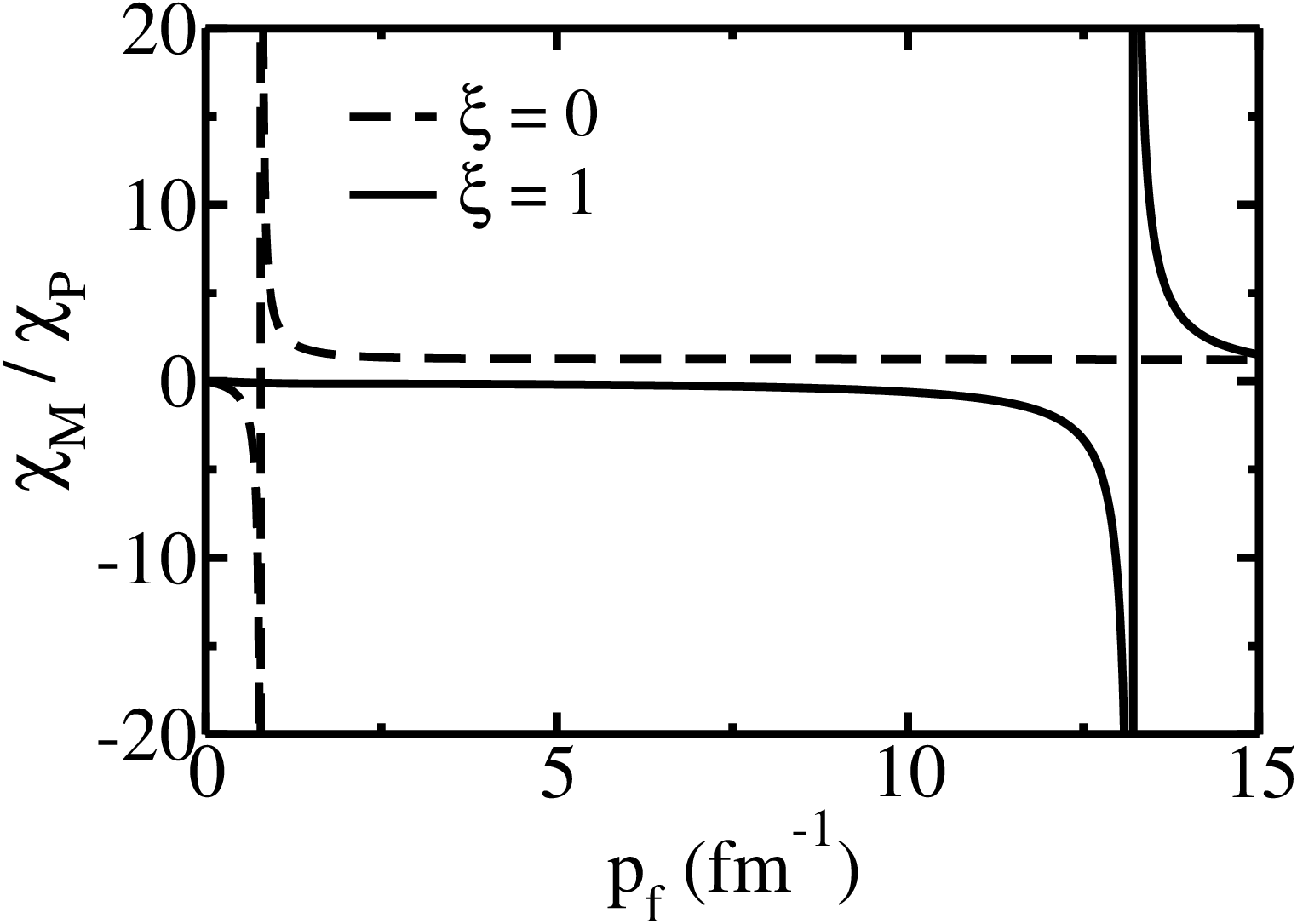}}
\caption{Magnetic susceptibility vs Fermi momentum using screened gluon mass for unpolarized and complete polarized quark matter.}
\label{chi_polun}
\end{center}
\end{figure}


In Fig.{\ref{chi_polun}}, the density dependence of magnetic susceptibility
both for unpolarized and polarized matter has been shown. We see that
the magnetic susceptibility diverges at some critical density which
increases with increasing $\xi$. It is apparent from the figure that, 
if the critical density for para-ferro phase transition becomes lower than 
the critical density for the magnetic transition, the latter cannot take
place. Thus, we conclude, that the magnetic transition depends on the critical 
density of para-ferro phase transition.

\vskip 0.2in

\begin{figure}[htb]
\begin{center}
\resizebox{8.5cm}{7.0cm}{\includegraphics[]{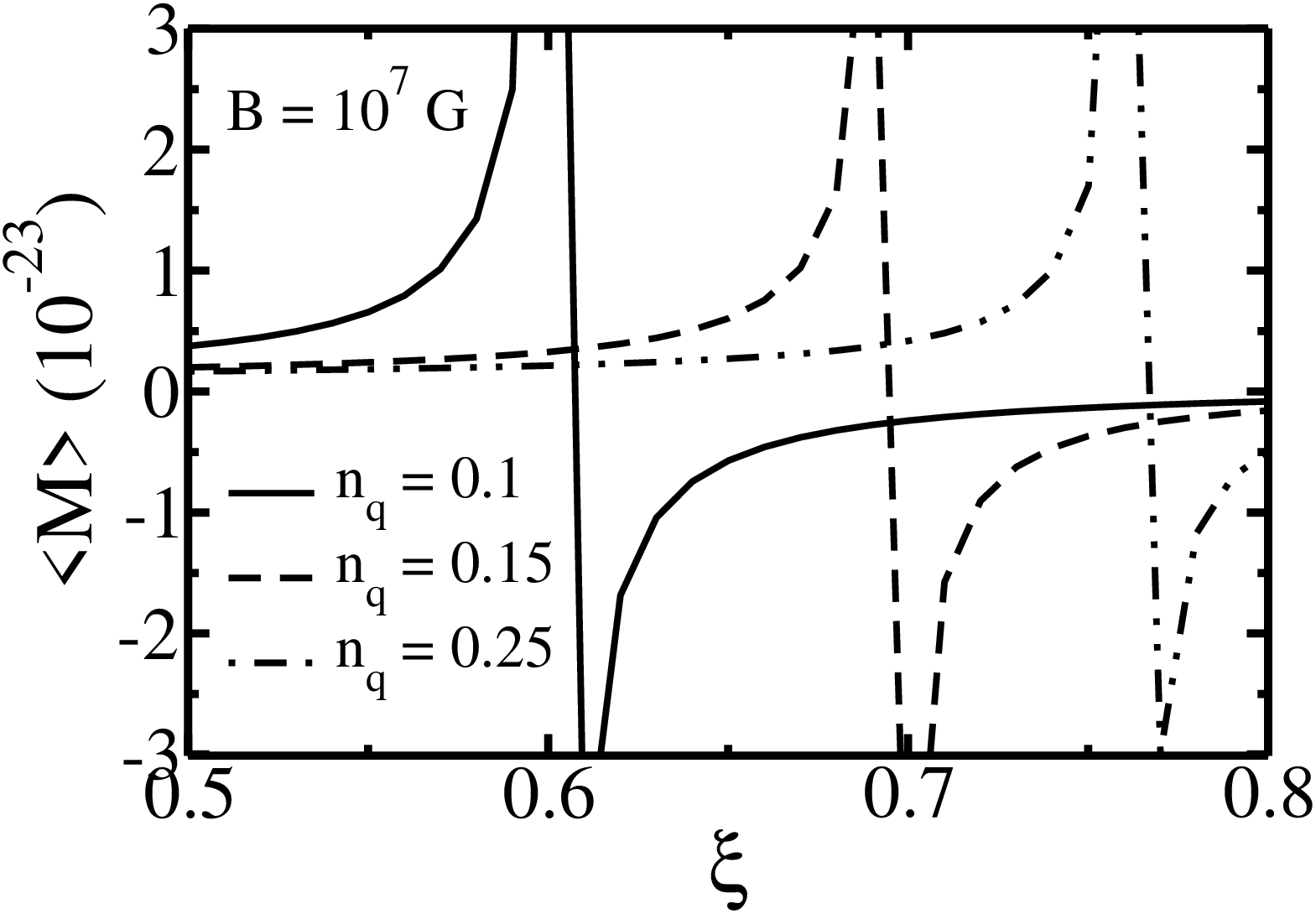}}
\caption{Variation of magnetization with $\xi$ for a magnetic field $B=10^7$ G.
Solid, dashed and dash-dotted lines represent $0.1 \rm{fm^{-3}}$, 
$0.15 \rm{fm^{-3}}$ and $0.25 \rm{fm^{-3}}$ respectively.}
\label{magnetization}
\end{center}
\end{figure}


In Fig.{\ref{magnetization}}, we show $\xi$  dependence of the
magnetization for various densities. Note that the
divergences appear at higher $\xi$ for larger density. Here 
the magnetic dipole moments of the quarks are taken to be: 
$\mu_{u}=1.852 \mu_{N}$,
$\mu_{d}=-0.972\mu_{N}$ and $\mu_{s}=-0.581\mu_{N}$, where 
the nuclear magneton $\mu_{N}=3.152 \times 10^{-14}$ MeV/ Tesla 
\cite{some96}. 

\section{Summary and conclusion}

In this work, we calculate dimensionless LPs $F_{0,1}^{sym}$ and 
$F_{0,1}^{asym}$ for dense quark matter. These are then used to 
calculate magnetic susceptibility and magnetization of
degenerate quark matter and the results are found to be consistent with
previous calculations in the appropriate limits. The qualitative
behavior of the FLPs as a function of density is also found to be
very similar to those of nuclear matter having isospin asymmetry.

We observe that $\chi_M$ is free of all the infrared divergences
even in the massless gluon limit. It is, however, numerically sensitive 
to the Debye mass. It is shown that the critical density for the magnetic
transition in polarized matter is higher than that of the
unpolarized one. The divergence and sign change of the magnetic 
susceptibility signal the magnetic instability of the ferromagnetic phase. 

\newpage
{\bf Acknowledgments}\\

The authors would like to thank S.Mallik and P.Roy for their critical reading 
of the manuscript and T.Tatsumi for his useful suggestions.

\end{document}